\documentclass[preprint,showpacs,preprintnumbers,amsmath,amssymb,natbib]{revtex4}

\usepackage{graphicx}% Include figure files
\usepackage{epsfig}		
\usepackage{dcolumn}% Align table columns on decimal point
\usepackage{bm}% bold math
\def\0{\mbox{\tiny $0$}}
\def\1{\mbox{\tiny $1$}}
\def\2{\mbox{\tiny $2$}}
\def\3{\mbox{\tiny $3$}}
\def\4{\mbox{\tiny $4$}}
\def\5{\mbox{\tiny $5$}}
\def\6{\mbox{\tiny $6$}}
\def\7{\mbox{\tiny $7$}}
\def\8{\mbox{\tiny $8$}}
\def\9{\mbox{\tiny $9$}}

\def\f14{\mbox{\tiny $\frac{1}{4}$}}

\def\ii{\mbox{\tiny $i$}}

\def\s{\mbox{\tiny $s$}}

\def\j{\mbox{\tiny $j$}}
\def\mi{\mbox{\tiny $-$}}

\def\pl{\mbox{\tiny $+$}}
\def\al{\mbox{\tiny $\alpha$}}

\def\bb#1{\mbox{\footnotesize $(#1)$}}
\begin{document}

%\preprint{DF/IST-10.2007}
%\preprint{December 2007}
%\title{Decoupling mass varying dark matter from an effective GCG through dark energy cosmon fields}
\title{Mass varying dark matter in effective GCG scenarios}

\author{A. E. Bernardini}
\affiliation{Departamento de F\'{\i}sica, Universidade Federal de S\~ao Carlos, PO Box 676, 13565-905, S\~ao Carlos, SP, Brasil}
\email{alexeb@ufscar.br, alexeb@ifi.unicamp.br}

\date{\today}% It is always \today, today,

\begin{abstract}
A unified treatment of mass varying dark matter coupled to cosmon-{\em like} dark energy is shown to result in {\em effective} generalized Chaplygin gas (GCG) scenarios.
The mass varying mechanism is treated as a cosmon field inherent effect.
Coupling dark matter with dark energy allows for reproducing the conditions for the present cosmic acceleration and for recovering the stability resulted from a positive squared speed of sound $c_{s}^{\2}$, as in the GCG scenario.
The scalar field mediates the nontrivial coupling between the dark matter sector and the sector responsible for the accelerated expansion of the universe.
The equation of state of perturbations is the same as that of the background cosmology so that all the effective results from the GCG paradigm are maintained.
Our results suggest the mass varying mechanism, when obtained from an exactly soluble field theory, as the right responsible for the stability issue and for the cosmic acceleration of the universe.
\end{abstract}

\date{\today}
\maketitle
\section{Introduction}
The ultimate nature of the dark sector of the universe is the most relevant issue related with the negative pressure component required to understand why and how the universe is undergoing a period of accelerated expansion \cite{Zla98,Wan99,Ste99,Bar99,Ber00}.
A natural and simplistic explanation for this is obtained in terms of a tiny positive cosmological constant introduced in the Einstein's equation for the universe.
Since the cosmological constant has a magnitude completely different from that predicted by theoretical arguments, and it is often confronted with conceptual problems, physicists have been compelled to consider other explanations for that \cite{Ame02,Kam02,Bil02,Ber02,Cal03,Mot04,Bro06A}.

Motivated by the high energy physics, an alternative for obtaining a negative pressure equation of state considers that the dark energy can be attributed to the dynamics of a scalar field $\phi$ which realizes the present cosmic acceleration by evolving slowly down its potential $V\bb{\phi}$ \cite{Pee87,Rat87}.
These models assume that the vacuum energy can vary \cite{Bro33}.
Following theoretical as well as phenomenological arguments, several possibilities have been proposed, such as $k$-essence \cite{Chi00,Arm01}, phantom energy \cite{Sch01,Car03}, cosmon fields \cite{Wet87}, and also several types of modifications of gravity \cite{Def02,Car04,Ama06}.

One of the most challenging proposals concerns mass varying particles \cite{Hun00,Gu03,Far04,Bja08} coupled to the dark energy through a dynamical mass dependence on a light scalar field which drives the dark energy evolution in a kind of unified cosmological fluid.
The idea in the well-known mass varying mechanism \cite{Far04,Pec05,Bro06A,Bja08} is to introduce a coupling between a relic particle and the scalar field whose effective potential changes as a function of the relic particle density.
This coupled fluid is either interpreted as dark energy plus neutrinos, or as dark energy plus dark matter \cite{Ber08A,Ber08B}.
Such theories can possess an adiabatic regime in which the scalar field always sits at the minimum of its effective potential, which is set by the local mass varying particle density.
The relic particle mass is consequently generated from the vacuum expectation value of the scalar field and becomes linked to its dynamics by $m\bb{\phi}$.

A scenario which congregate dark energy and some kind of mass varying dark matter in a unified negative pressure fluid can explain the origin of the cosmic acceleration.
In particular, any background cosmological fluid with an effective behaviour as that of the generalized Chaplygin gas (GCG) \cite{Kam02,Bil02,Ber02} naturally offers this possibility.
The GCG is particularly relevant in respect with other cosmological models as it is shown to be consistent with the observational constraints from CMB \cite{Ber03}, supernova \cite{Sup1,Ber04,Ber05}, gravitational lensing surveys \cite{Ber03B}, and gamma ray bursts \cite{Ber06B}.
Moreover, it has been shown that the GCG model can be accommodated within the standard structure formation mechanism \cite{Kam02,Ber02,Ber04}.

In the scope of finding a natural explanation for the cosmic acceleration and the corresponding adequation to stability conditions for a background cosmological fluid, our purpose is to demonstrate that the GCG just corresponds to an effective description of a coupled fluid composed by dark energy with equation of state given by $p\bb{\phi} = -\rho\bb{\phi}$ and a cold dark matter (CDM) with a dynamical mass driven by the scalar field $\phi$.
Once one has consistently obtained the mass dependence on $\phi$, which is model dependent, it can be noticed that the cosmological evolution of the composed fluid is governed by the same dynamics prescribed by cosmon field equations.
It suggests that the mass varying mechanism embedded into the cosmon-{\em like} dynamics reproduces the effective behaviour of the GCG.
In addition, coupling dark matter with dark energy by means of a dynamical mass driven by such a scalar field allows for reproducing the conditions for the present cosmic acceleration and for recovering the stability prescribed by a positive squared speed of sound $c_{s}^{\2}$.
At least implicitly, it leads to the conclusion that the dynamical mass behaviour is the main agent of the stability issue and of the cosmic acceleration of the universe.

At our approach, the dark matter is approximated by a degenerate fermion gas (DFG).
In order to introduce the mass varying behaviour, we analyze the consequences of coupling it with and underlying dark energy scalar field driven by a cosmon-{\em type} equation of motion.
We discuss all the relevant constraints on this in section II.
In section III, we obtain the energy density and the equation of state for the unified fluid and compare them with the corresponding quantities for the GCG.
In section IV, we discuss the stability issue and the accelerated expansion of the universe in the framework here proposed.
The pertinent comparisons with a GCG scenario are evaluated.
We draw our conclusions in section V by summarizing our findings and discussing their implications.

\section{Mass varying mechanism for a DFG coupled to cosmon-{\em like} scalar fields}

To understand how the mass varying mechanism takes place for different particle species, it is convenient to describe the corresponding particle density, energy density and pressure as functionals of a statistical distribution.
This counts the number of particles in a given region around a point of the phase space defined by the conjugate coordinates: momentum, {\boldmath$\beta$}, and position, {\boldmath$x$}.
The statistical distribution can be defined by a function $f\bb{q}$ in terms of a comoving variable, $q =  a\,|\mbox{\boldmath$\beta$}|$, where $a$ is the scale factor (cosmological radius) for the flat FRW universe, for which the metrics is given by $ds^{\2} = dt^{\2} - a^{\2}\bb{t}\delta_{\ii\j}dx^{\ii}dx^{\j}$.
In the flat FRW scenario, the corresponding particle density, energy density and pressure are thus given by
\begin{eqnarray}
n\bb{a} &=&\frac{1}{\pi^{\2}\,a^{\3}}
\int_{_{0}}^{^{\infty}}{\hspace{-0.3cm}dq\,q^{\2}\ \hspace{-0.1cm}f\bb{q}},\nonumber\\
\rho_m\bb{a, \phi} &=&\frac{1}{\pi^{\2}\,a^{\4}}
\int_{_{0}}^{^{\infty}}{\hspace{-0.3cm}dq\,q^{\2}\, \left(q^{\2}+ m^{\2}\bb{\phi}\,a^{\2}\right)^{\1/\2}\hspace{-0.1cm}f\bb{q}},\\
p_m\bb{a, \phi} &=&\frac{1}{3\pi^{\2}\,a^{\4}}\int_{_{0}}^{^{\infty}}{\hspace{-0.3cm}dq\,q^{\4}\, \left(q^{\2}+ m^{\2}\bb{\phi}\,a^{\2}\right)^{\mi\1/\2}\hspace{-0.1cm} f\bb{q}}.~~~~ \nonumber
\label{gcg01}
\end{eqnarray}
where the last two can be depicted from the Einstein's energy-momentum tensor \cite{Dod05}.
For the case where $f\bb{q}$ is a Fermi-Dirac distribution function, it can be written as
\begin{equation}
f\bb{q}= \left\{\exp{\left[(q - q_F)/T_{\0}\right]} + 1\right\}^{\mi\1}, \nonumber
\end{equation}
where $T_{\0}$ is the relic particle background temperature at present.
In the limit where $T_{\0}$ tends to $0$, it becomes a step function that yields an elementary integral for the above equations, with the upper limit equal to the Fermi momentum here written as $q_{F} = a \,\beta\bb{a}$.
It results in the equations for a DFG \cite{ZelXX}.
The equation of state can be expressed in terms of elementary functions of $\beta \equiv \beta\bb{a}$ and $m \equiv m\bb{\phi\bb{a}}$,
\begin{eqnarray}
n\bb{a} &=& \frac{1}{3 \pi^{\2}} \beta^{\3},\nonumber\\
\rho_m\bb{a} &=& \frac{1}{8 \pi^{\2}}
\left[\beta(2 \beta^{\2} + m^{\2})\sqrt{\beta^{\2} + m^{\2}} -
\mbox{arc}\sinh{\left(\beta/m\right)}\right],\\
p_m\bb{a} &=& \frac{1}{8 \pi^{\2}}
\left[\beta (\frac{2}{3} \beta^{\2} - m^{\2})\sqrt{\beta^{\2} + m^{\2}} + \mbox{arc}\sinh{\left(\beta/m\right)}\right].\nonumber
\label{gcg01B}
\end{eqnarray}
One can notice that the DFG approach is useful for parameterizing the transition between ultra-relativistic (UR) and non-relativistic (NR) thermodynamic regimes.
It is not mandatory for connecting the mass varying scenario with the GCG scenario.

Simple mathematical manipulations allow one to easily demonstrate that
\begin{equation}
n\bb{a} \frac{\partial \rho_m\bb{a}}{\partial n\bb{a}} = (\rho_m\bb{a} + p_m\bb{a}),
\label{gcg02B}
\end{equation}
and
\begin{equation}
m\bb{a} \frac{\partial \rho_m\bb{a}}{\partial m\bb{a}} = (\rho_m\bb{a} - 3 p_m\bb{a}),
\label{gcg02}
\end{equation}

Noticing that the explicit dependence of $\rho_m$ on $a$ is intermediated by $\beta\bb{a}$ and $m\bb{a} \equiv m\bb{\phi\bb{a}}$, one can take the derivative of the energy density with respect to time in order to obtain
\begin{eqnarray}
\dot{\rho}_m &=& \dot{\beta}\bb{a} \frac{\partial \rho_m\bb{a}}{\partial \beta\bb{a}}  + \dot{m}\bb{a} \frac{\partial \rho_m\bb{a}}{\partial m\bb{a}}\nonumber\\
             &=& \dot{n}\bb{a} \frac{\partial \rho_m\bb{a}}{\partial n \bb{a}}  + \dot{m}\bb{a} \frac{\partial \rho_m\bb{a}}{\partial m\bb{a}}\nonumber\\
             &=& - 3\frac{\dot{a}}{a} n\bb{a} \frac{\partial \rho_m\bb{a}}{\partial n \bb{a}}  + \dot{\phi}\frac{\mbox{d} m}{\mbox{d} \phi}\frac{\partial \rho_m\bb{a}}{\partial m\bb{a}},
\label{gcg03BB}
\end{eqnarray}
where the {\em overdot} denotes differentiation with respect to time ($^{\cdot}\, \equiv\, d/dt$).
The substitution of Eqs.~(\ref{gcg02B}-\ref{gcg02}) into the above equation results in the energy conservation equation given by
\begin{equation}
\dot{\rho}_m + 3 H (\rho_m + p_m) - \dot{\phi}\frac{\mbox{d} m}{\mbox{d} \phi} (\rho_m  - 3 p_m) = 0,
\label{gcg03}
\end{equation}
where $H = \dot{a}/{a}$ is the expansion rate of the universe.
If one performs the derivative with respect to $a$ directly from $\rho_m$ in the form given in Eq.~(\ref{gcg01}), the same result can be obtained.
The coupling between relic particles and the scalar field as described by Eq.~(\ref{gcg03}) are effective just for NR fluids.
Since the strength of the coupling is suppressed by the relativistic increase of pressure ($\rho\sim 3 p$),
as long as particles become relativistic ($T\bb{a} = T_{\0}/a >> m\bb{\phi\bb{a}}$) the matter fluid and the scalar field fluid tend to decouple and evolve adiabatically.
The mass varying mechanism expressed by Eq.~(\ref{gcg03}) translates the dependence of $m$ on $\phi$ into a dynamical behaviour.
In particular, for a DFG, the consistent analytical transition between UR and NR regimes and their effects on coupling dark matter ($m$) and dark energy ($\phi$) are evident from Eq.~(\ref{gcg03}).
The mass thus depends on the value of a slowly varying classical scalar field \cite{Wet94, Bea08} which evolves like a {\em cosmon} field.
The cosmon-{\em type} equation of motion for the scalar field $\phi$ is given by
\begin{equation}
\ddot{\phi} + 3 H \dot{\phi} + \frac{\mbox{d} V\bb{\phi}}{\mbox{d} \phi} =  Q\bb{\phi}.
\label{gcg04}
\end{equation}
where, in the mass varying scenario, one identifies $Q\bb{\phi}$ as $ - (\mbox{d} m/\mbox{d} \phi)/(\partial \rho_m/\partial m)$.
The corresponding equation for energy conservation can be written as
\begin{equation}
\dot{\rho_{\phi}} + 3 H (\rho_{\phi} + p_{\phi}) + \dot{\phi}\frac{\mbox{d} m}{\mbox{d} \phi} \frac{\partial \rho_m}{\partial m} = 0.
\label{gcg05}
\end{equation}
which, when added to Eq.~(\ref{gcg03}), results in the equation  for a unified fluid $(\rho, p)$ with a dark energy component and a mass varying dark matter component,
\begin{equation}
\dot{\rho} + 3 H (\rho + p) = 0,
\label{gcg06}
\end{equation}
where $\rho = \rho_{\phi} + \rho_m$ and $p = p_{\phi} + p_m$.

As we shall notice in the following, this unified fluid corresponds to an effective description of the universe parameterized by a GCG equation of state.

\section{Decoupling mass varying dark matter from the effective GCG}

Irrespective of its origin, several studies yield convincing evidences that the GCG scenario is phenomenologically consistent with the accelerated expansion of the universe.
This scenario is introduced by means of an exotic equation of state \cite{Ber02,Kam02,Ber03} given by
\begin{equation}
p = - A_{\s} \left(\frac{\rho_{\0}}{\rho}\right)^{\al},
\label{gcg20}
\end{equation}
which can be obtained from a generalized Born-Infeld action \cite{Ber02}.
The constants $A_{\s}$ and $\alpha$ are positive and $0 < \alpha \leq 1$.
Of course, $\alpha = 0$ corresponds to the $\Lambda$CDM model and we are assuming that the GCG model has an underlying scalar field, actually real \cite{Kam02,Ber04} or complex \cite{Bil02,Ber02}.
The case $\alpha = 1$ corresponds to the equation of state of the Chaplygin gas scenario \cite{Kam02} and is already ruled out by data \cite{Ber03}.
Notice that for $A_s =0$, GCG behaves always as matter whereas for $A_{\s} =1$, it behaves always as a cosmological constant.
Hence to use it as a unified candidate for dark matter and dark energy one has to exclude these two possibilities so that $A_s$ must lie in the range $0 < A_{\s} < 1$.

Inserting the above equation of state into the unperturbed energy conservation Eq.~(\ref{gcg06}), one obtains through a straightforward integration \cite{Kam02,Ber02}
\begin{equation}
\rho = \rho_{\0} \left[A_{\s} + \frac{(1-A_{\s})}{a^{\3(\1+\alpha)}}\right]^{\1/(\1 \pl \al)},
\label{gcg21}
\end{equation}
and
\begin{equation}
p = - A_{\s} \rho_{\0} \left[A_{\s} + \frac{(1-A_{\s})}{a^{\3(\1+\alpha)}}\right]^{-\al/(\1 \pl \al)}.
\label{gcg22}
\end{equation}

One of the most striking features of the GCG fluid is that its energy density interpolates between a dust dominated phase, $\rho \propto a^{-\3}$, in the past, and a de-Sitter phase, $\rho = -p$, at late times.
This property makes the GCG model an interesting candidate for the unification of dark matter and dark energy.
Indeed, it can be shown that the GCG model admits inhomogeneities and that, in particular, in the context of the Zeldovich approximation, these evolve in a qualitatively similar way as in the $\Lambda$CDM model \cite{Ber02}.
Furthermore, this evolution is controlled by the model parameters, $\alpha$ and $A_{\s}$.

Assuming the canonical parametrization of $\rho$ and $p$ in terms of a scalar field $\phi$,
\begin{eqnarray}
\rho &=& \frac{1}{2}\dot{\phi}^{\2} + V,\nonumber\\
p    &=& \frac{1}{2}\dot{\phi}^{\2} - V,
\label{pap01}
\end{eqnarray}
allows for obtaining the effective dependence of the scalar field $\phi$ on the scale factor, $a$, and explicit expressions for $\rho$, $p$ and $V$ in terms of $\phi$.

Following Ref.~\cite{Ber04}, one can obtain through Eq.~(\ref{gcg05}) the field dependence on $a$,
\begin{equation}
\dot{\phi}^{\2}\bb{a} = \frac{\rho_{\0}(1 - A_{\s})}{a^{\3(\al\pl\1)}}
\left[A_{\s} + \frac{(1-A_{\s})}{a^{\3(\al\pl\1)}}\right]^{-\al/(\al \pl \1)},
\label{pap02}
\end{equation}
and assuming a flat evolving universe described by the Friedmann equation $H^{\2} = \rho$ (with $H$ in units of $H_{\0}$ and $\rho$ in units of $\rho_{\mbox{\tiny Crit}} = 3 H^{\2}_{\0}/ 8 \pi G)$, one obtains
\begin{equation}
\phi\bb{a} = - \frac{1}{3(\alpha + 1)}\ln{\left[\frac{\sqrt{1 - A_{\s}(1 - a^{\3(\al \pl \1)})} - \sqrt{1 - A_{\s}}}{\sqrt{1 - A_{\s}(1 - a^{\3(\al \pl \1)})} + \sqrt{1 - A_{\s}}}\right]},
\label{pap03}
\end{equation}
where it is assumed that
\begin{equation}
\phi_{\0} = \phi\bb{a_{\0} = 1} = - \frac{1}{3(\alpha + 1)}\ln{\left[\frac{1 - \sqrt{1 - A_{\s}}}{1 + \sqrt{1 - A_{\s}}}\right]}.
\label{pap04}
\end{equation}
One then readily finds the scalar field potential,
\begin{equation}
V\bb{\phi} = \frac{1}{2}A_{\s}^{\frac{\1}{\1 \pl \al}}\rho_{\0}\left\{
\left[\cosh{\left(3\bb{\alpha + 1} \phi/2\right)}\right]^{\frac{\2}{\al \pl \1}}
+
\left[\cosh{\left(3\bb{\alpha + 1} \phi/2\right)}\right]^{-\frac{\2\al}{\al \pl \1}}
\right\}.
\label{pap05}
\end{equation}
If one supposes that energy density, $\rho$, may be decomposed into a mass varying CDM component, $\rho_{m}$, and a dark energy component, $\rho_{\phi}$, connected by the scalar field equations (\ref{gcg04})-(\ref{gcg05}), the equation of state (\ref{gcg20}) is just assumed as an effective description of the cosmological background fluid of the universe.
Since the CDM pressure, $p_m$, is null, the dark energy component of pressure, $p_{\phi}$, results in the GCG pressure, $p = p_{\phi}$.
Assuming that dark energy obeys a de-Sitter phase equation of state, that is, $\rho_{\phi}\bb{\phi} = - p_{\phi}\bb{\phi}$, the dark energy density can be parameterized by a generic quintessence potential, $\rho_{\phi}\bb{\phi} = U\bb{\phi}$, since its kinetic component has to be null for a canonical formulation.
It results in $U\bb{\phi} = - p_{\phi}\bb{\phi} = p$, where $p$ is the GCG pressure given by Eq.~(\ref{gcg22}).
By substituting the result of Eq.(\ref{pap03}) into the Eq.(\ref{gcg22}), and observing that $H^{\2} = \rho$, with $\rho$ given by Eq.(\ref{gcg21}), it is possible to rewrite the GCG pressure, $p$, in terms of $\phi$.
It results in the following analytical expression for $U\bb{\phi}$,
\begin{equation}
U\bb{\phi} = \rho_{\phi}\bb{\phi} = - p_{\phi}\bb{\phi} = \left[A_{s}\cosh{\left(\frac{3\bb{\alpha + 1}\phi}{2}\right)}\right]^{\frac{\2 \al}{1 \pl \al}},
\label{pap08}
\end{equation}
which is consistent with the result for $V\bb{\phi} = (1/2)(\rho\bb{\phi} - p\bb{\phi})$ from Eq.~(\ref{pap05}).
Since $\rho_{\phi}\bb{\phi} + p_{\phi}\bb{\phi} = 0$, the Eq.~(\ref{gcg05}) is thus reduced to
\begin{equation}
\frac{\mbox{d} U\bb{\phi}}{\mbox{d}{\phi}} +
\frac{\partial \rho_{m}}{\partial m} \frac{\mbox{d} m\bb{\phi}}{\mbox{d}\phi} = 0,
\label{pap09}
\end{equation}
and the problem is then reduced to finding a relation between the scalar potential $U\bb{\phi}$ and the variable mass $m\bb{\phi}$.
From the above equation, the effective potential governing the evolution of the scalar field is naturally decomposed into a sum of two terms, one arising from the original quintessence potential $U\bb{\phi}$, and other from the dynamical mass $m\bb{\phi}$.
For appropriate choices of potentials and coupling functions satisfying Eq.~(\ref{pap09}), the competition between these terms leads to a minimum of the effective potential.
For {\em quasi}-static regimes, it is possible to adiabatically track the position of this minimum, in a kind of stationary condition.
The timescale for $\phi$ to adjust itself to the dynamically modified minimum of the effective potential may be short compared to the timescale over which the background density is changing.
In the adiabatic regime, the matter and scalar field are tightly coupled together and evolve as one effective fluid.
At our approach, once we have assumed the dark energy equation of state as $p_{\phi} = - \rho_{\phi}$, the stationary condition is a natural issue that emerges without any additional constraint on cosmon-{\em type} equations.
In the GCG cosmological scenario, the effective fluid description is valid for the background cosmology and for linear perturbations.
The equation of state of perturbations is the same as that of the background cosmology where all the effective results of the GCG paradigm are maintained.

The Eq.~(\ref{pap08}) leads to $\rho + p = \rho_{m} + p_{m}$ which, in the CDM limit, gives
\begin{equation}
\rho\bb{a} + p\bb{a} = m\bb{a} \, n\bb{a} + p_{m} (\equiv 0) = \frac{1}{3\pi^{\2}}\,m\bb{a}\, \beta^{\3}\bb{a}.
\label{pap10}
\end{equation}
Since the dependence of $m$ on $a$ is exclusively intermediated by $\phi\bb{a}$, i. e. $m\bb{a} \equiv m\bb{\phi\bb{a}}$,
from Eqs.~(\ref{gcg21}), (\ref{gcg22}) and (\ref{pap03}), after some mathematical manipulations, one obtains
\begin{equation}
m\bb{\phi} = m_{\0} \left[\frac{\tanh{\left(3\bb{\alpha + 1}\frac{\phi}{2}\right)}}{\tanh{\left(3\bb{\alpha + 1}\frac{\phi_{\0}}{2}\right)}}\right]^{\frac{ \2 \al}{1 \pl \al}}
\label{pap11}
\end{equation}
which is consistent with Eq.~(\ref{pap09}) once $n \propto a^{\mi\3}$.
One can thus infer that the adequacy to the adiabatic regime is left to the mass varying mechanism which drives the cosmological evolution of the dark matter component.

\begin{figure}
\vspace{-2 cm}
\centerline{\psfig{file= 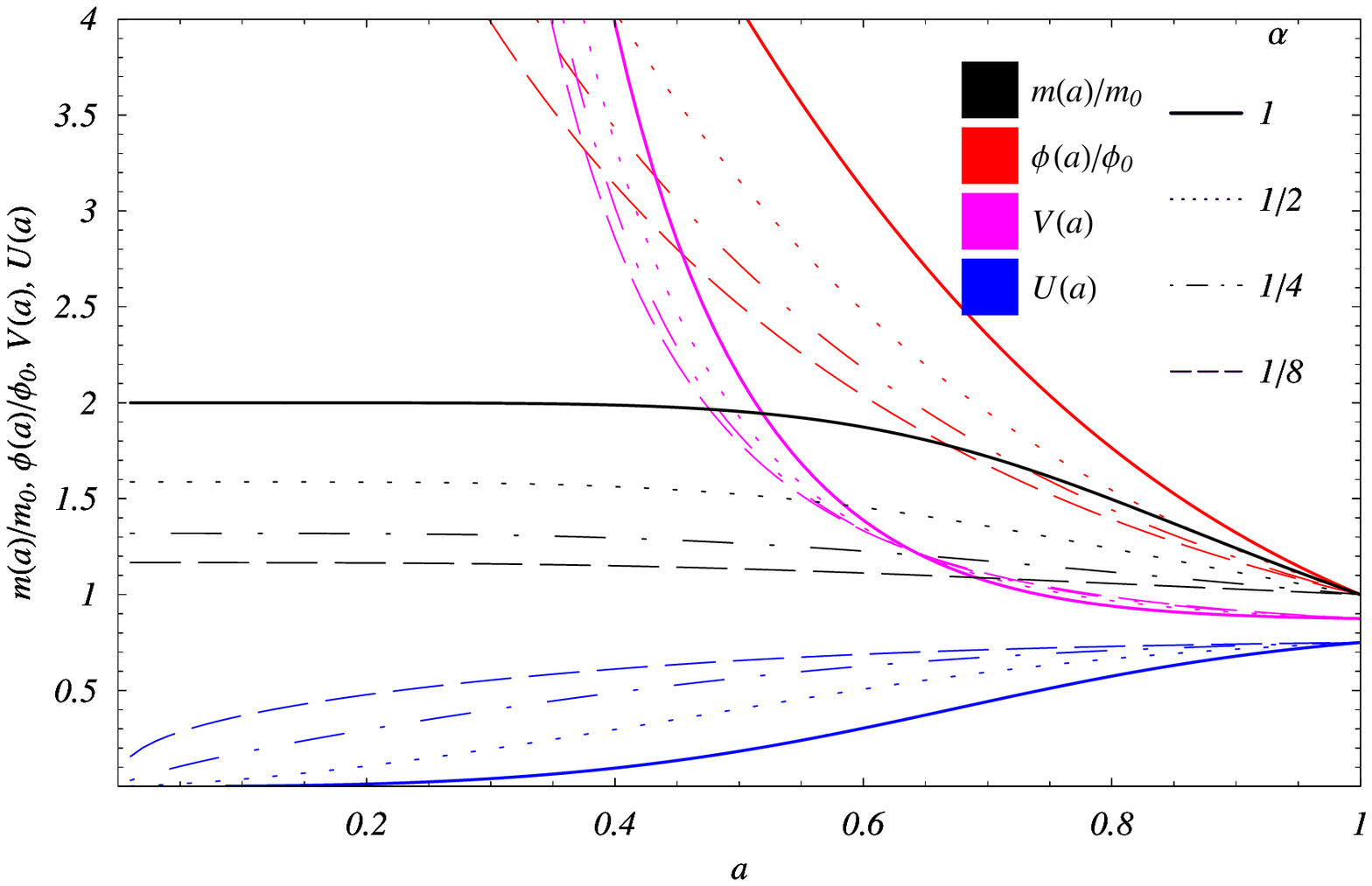,width=13.5 cm}}
\vspace{-3 cm}
\centerline{\psfig{file= 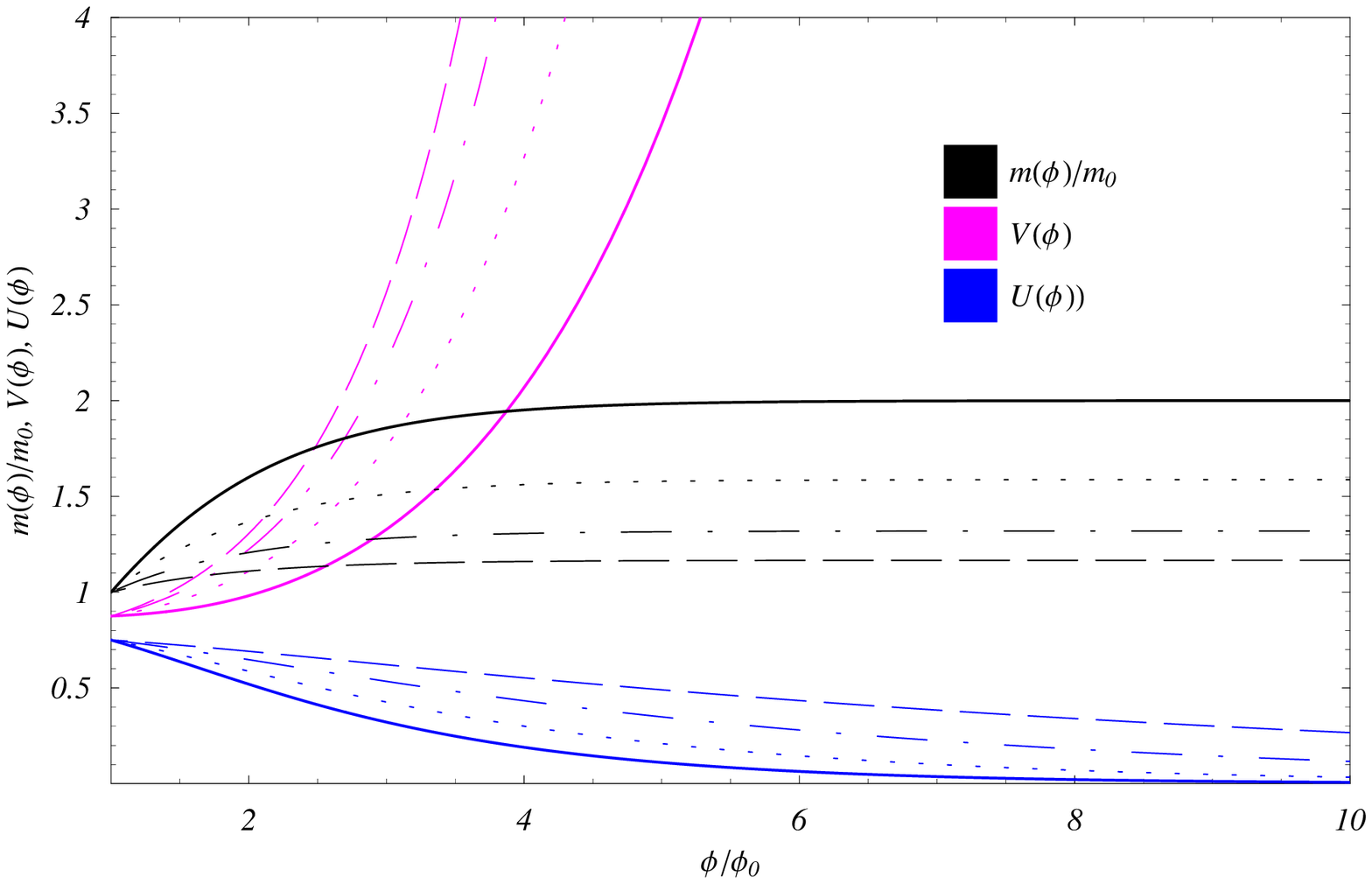,width=13 cm}}
\vspace{-3 cm}
\caption{\small The cosmological evolution of $m\bb{a}$ and $U\bb{a}$ correlated with that of $\phi\bb{a}$ and $V\bb{a}$ of the GCG as function of the scale factor (first plot) and the corresponding dependence on $\phi$ (second plot).
We are considering the GCG scenarios with $A_{\s}= 3/4$ and $\alpha = 1,\,1/2,\,1/4,\,1/8$.}
\label{Fpap-01}
\end{figure}

\begin{figure}
\vspace{-2 cm}
\centerline{\psfig{file= 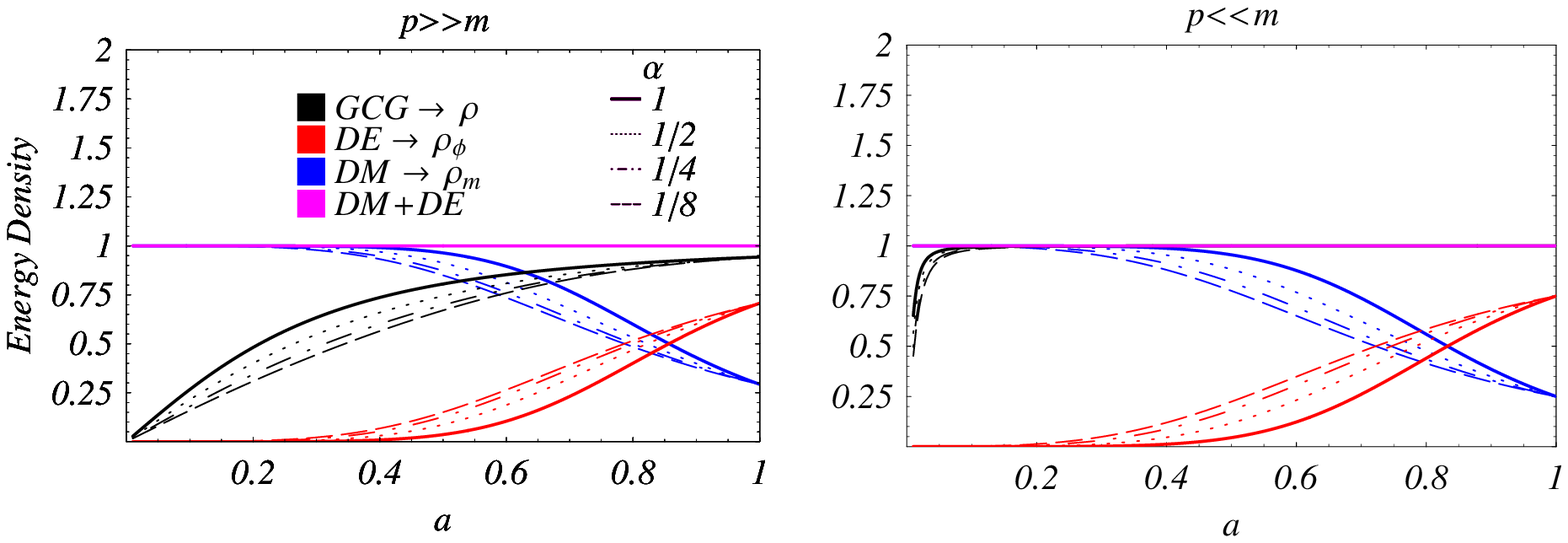,width= 14 cm}}
\vspace{-3 cm}
\centerline{\psfig{file= 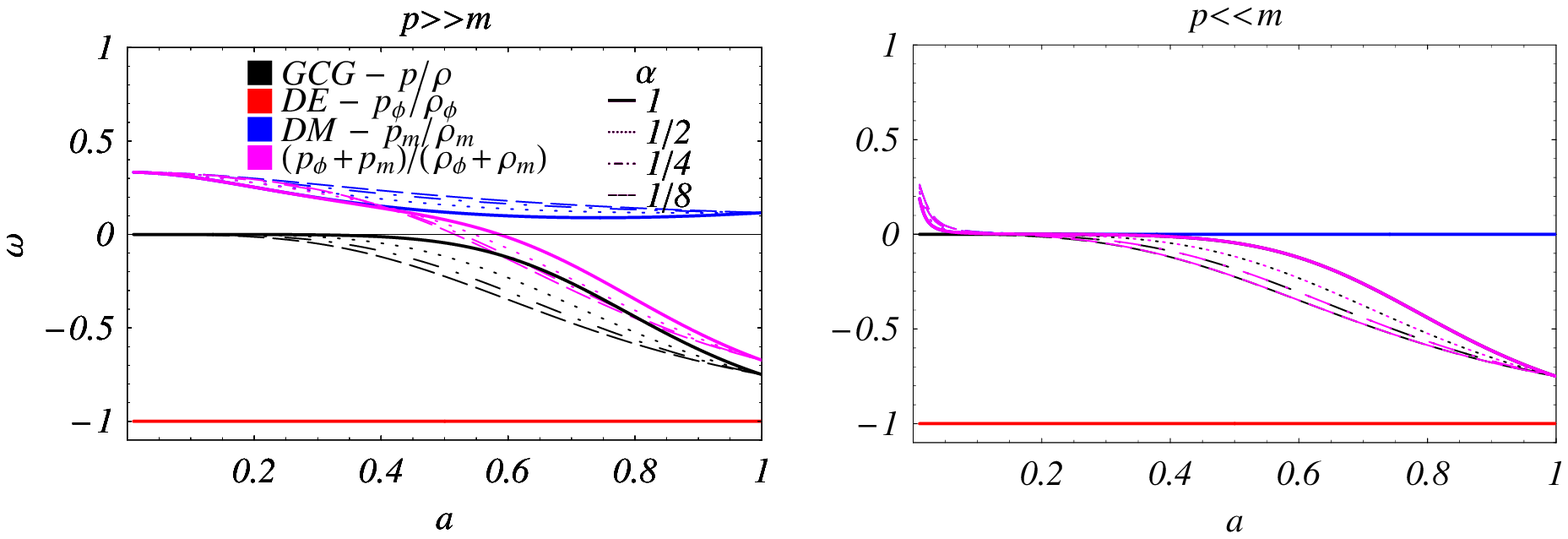,width= 14 cm}}
\vspace{-4 cm}
\caption{\small Energy densities, $\rho$, and equations of state, $\omega = p /\rho$, as function of the scale factor for the effective GCG fluid and its components: mass varying dark matter $\rho_{m}$ and cosmon-{\em like} dark energy $\rho_{\phi}$.
The results are compared with those of real GCG scenarios with $A_{\s}= 3/4$ and $\alpha = 1,\,1/2,\,1/4,\,1/8$.
Dark matter is assumed to be described by a DFG in relativistic ($p >> m$) and non-relativistic regimes ($p << m$) at present.}
\label{Fpap-02}
\end{figure}

To give the correct impression of the time evolution of the abovementioned dynamical quantities driven by $\phi$, in the Fig.~\ref{Fpap-01} we observe the behaviour of $m\bb{a}$ and $U\bb{a}$ in confront with $\phi\bb{a}$ and $V\bb{a}$ of the GCG.
In the Fig.~\ref{Fpap-02} we verify how the energy density $\rho$ and the corresponding equations of state $\omega$ for the unified fluid, $\rho_{m} + \rho_{\phi}$, which imitates the GCG, deviates from the right GCG scenario.
We assume that the mass varying dark matter behaves like a DFG in a relativistic regime (hot dark matter (HDM)) and in a non-relativistic regime (CDM).
For mass varying CDM coupled with dark energy with $p_{\phi} = -\rho_{\phi}$, the effective GCC leads to similar predictions for $\omega$, independently of the scale parameter $a$.
The same is not true for HDM which, in the DFG approach, when weakly coupled with dark energy, leads to the same behavior of the GCG just for late time values of $a$ ($a \sim 1$).

As one can observe, the mass varying mechanism allows for reconstituting the GCG scenario in terms of non-exotic primitive entities: CDM and dark energy.
Furthermore, as we shall notice, the dynamical mass expressed by Eq.~(\ref{pap11}) overpasses the problematic issue of stability.

It is also important to emphasize that, as in the case of the Chaplygin gas, where $\alpha = 1$, the GCG model admits a d-brane connection as its Lagrangian density corresponds to the Born-Infeld action plus some soft logarithmic corrections.
Space-time is shown to evolve from a phase that is initially dominated, in the absence of other degrees of freedom on the brane, by non-relativistic matter to a phase that is asymptotically De Sitter.
The Chaplygin gas reproduces such a behaviour.
In this context, the explicit dependence of the mass on a scalar field is relevant in suggesting that the mass varying mechanism, when obtained from an exactly soluble field theory, can be the right responsible for the cosmological dynamics.

\section{Stability and accelerated expansion}

Adiabatic instabilities in cosmological scenarios was predicted \cite{Afs05} in a context of a mass varying neutrino (MaVaN) model of dark energy.
The dynamical dark energy, in this approach, is obtained by coupling a light scalar field to neutrinos but not to dark matter.
Their consequent effects have been extensively discussed in the context of mass varying neutrinos, in which the light mass of the neutrino and the recent accelerative era are twinned together through a scalar field coupling.
In the adiabatic regime, these models faces catastrophic instabilities on small scales characterized by a negative squared speed of sound for the effectively coupled fluid.
Starting with a uniform fluid, such instabilities would give rise to exponential growth of small perturbations.
The natural interpretation of this is that the Universe becomes inhomogeneous with neutrino overdensities subject to nonlinear fluctuations \cite{Mot08} which eventually collapses into compact localized regions.

In opposition, in the usual treatment where dark matter are just coupled to dark energy, cosmic expansion together with the gravitational drag due to CDM have a major impact on the stability of the cosmological background fluid.
Usually, for a general fluid for which we know the equation of state, the dominant effect on the sound speed squared $c_{s}^{\2}$ arises from the dark sector component and not by the neutrino component.

\begin{figure}
\vspace{-2 cm}
\centerline{\psfig{file= 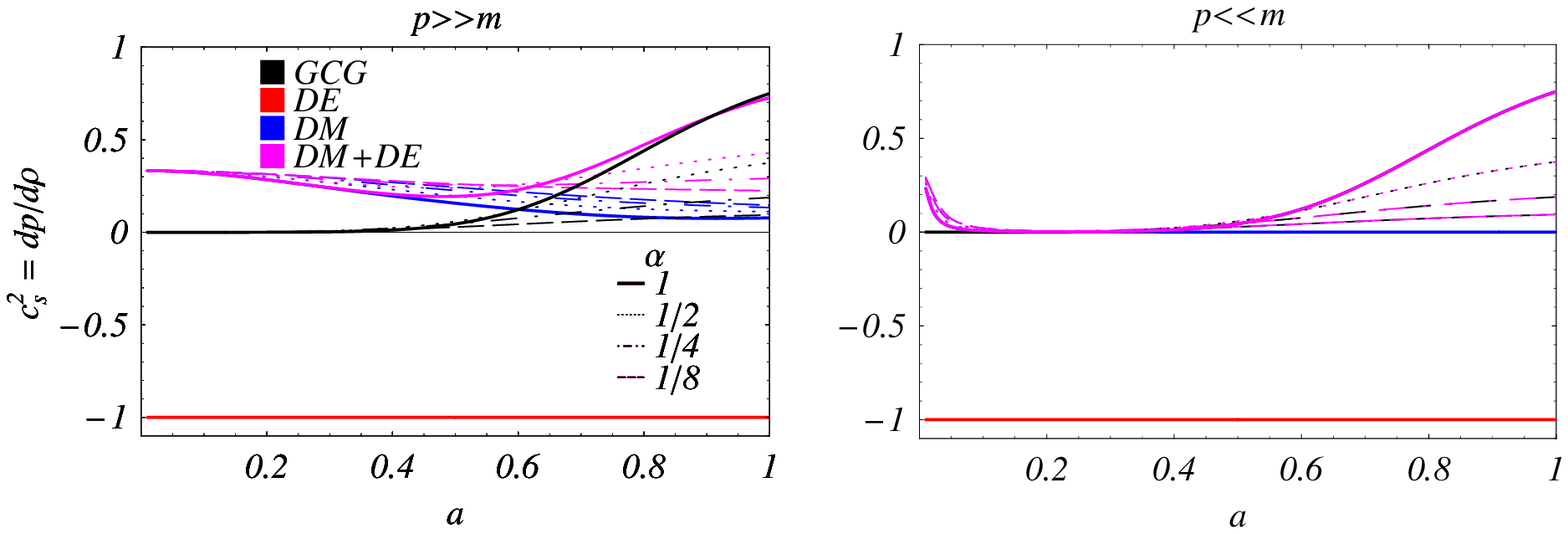,width= 14 cm}}
\vspace{-3 cm}
\centerline{\psfig{file= 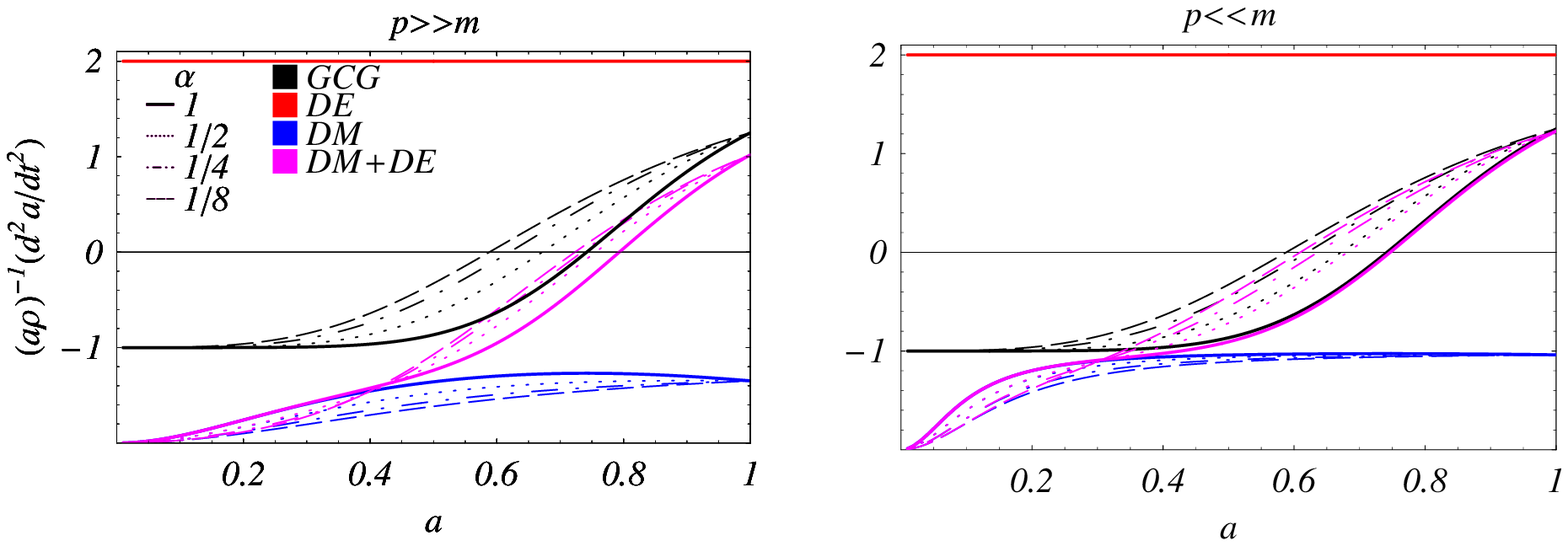,width= 14 cm}}
\vspace{-4 cm}
\caption{\small Squared speed of sound $c^{\2}_{s} = \mbox{d}p/\mbox{d}\rho$ and cosmic acceleration $\ddot{a}/(a\,
\rho)$ as function of the scale factor for the unified fluid decomposed into mass varying dark matter and cosmon-{\em like} dark energy components.
Again the results are compared with those of real GCG scenarios for $A_{\s}= 3/4$ and $\alpha = 1,\,1/2,\,1/4,\,1/8$, for mass varying dark matter described by a DFG in relativistic ($p >> m$) and non-relativistic regimes ($p << m$) at present.}
\label{Fpap-03}
\end{figure}

For the models where the stationary condition (cf. Eq.~(\ref{pap09})) implies a cosmological constant type equation of state, $ p_{\phi} = - \rho_{\phi}$, one obtains $c_{s}^{\2} = -1$ from the very start of the analysis.
The effective GCG is free from this inconsistency.
The coupling of the dark energy component with dynamical dark matter is responsible for removing such inconsistency by setting $c_{s}^{\2} \simeq \frac{d p_{\phi}}{d\rho_{\phi}} > 0$.
The exact behavior for dark energy plus mass varying dark matter fluid in correspondence with the GCG is exhibited in the Fig.~\ref{Fpap-03} for different GCG $\alpha$ parameters.
A previous analysis of the stability conditions for the GCG in terms of the squared speed of sound was introduced in Ref. \cite{Ber04}, from which positive $c_{s}^{\2}$ implies that $0 \leq \alpha \leq 1$.
These results are consistent with the accelerated expansion of the universe ruled by the dynamical mass of Eq.~(\ref{pap11}) which sets positive values for $(1 + 3 (p_{\phi} + p_m)/(\rho_{\phi} + \rho_m))$, as we can notice in the Fig.~\ref{Fpap-03}.

For CDM ($p << m$) the unified fluid reproduces the GCG scenario.
For HDM ($p >> m$), in spite of not reproducing the GCG, the conditions for stability and cosmic acceleration are maintained.
Fig.~\ref{Fpap-03} shows that the GCG can indeed be interpreted as the effective result for the coupling between mass varying dark matter and a kind of scalar field dark energy which is cosmologically driven by a $\Lambda$-type equation of state, $p_{\phi} = -\rho_{\phi}$.

\section{Conclusions}

The dynamics of the cosmology of mass varying dark matter coupled with dark energy dynamically driven by cosmon-{\em type} equations was studied without introducing specific quintessence potentials, but assuming that the cosmological background unified fluid presents an effective behaviour similar to that of the GCG.

We have comprehensively analyzed the stability characterized by a positive squared speed of sound and the cosmic acceleration conditions for such a dark matter coupled to dark energy fluid, that exists whenever such theories enter an adiabatic regime in which the scalar field faithfully tracks the minimum of the effective potential, and the coupling strength is strong compared to gravitational strength.
The matter and scalar field are tightly coupled together and evolve as one effective fluid.
The effective potential governing the evolution of the scalar field is decomposed into a sum of two terms, one arising from the original scalar field potential $U\bb{\phi}$, and the other from the dynamical mass of the dark matter.

The mass varying behaviour of the dark matter component was determined from the assumption of a kind of $\Lambda$-type dark energy dynamics embedded in an effective cosmological scenario which reproduces the cosmological effects of the GCG.
It is equivalent to decoupling mass varying dark matter from the effective GCG concomitantly with assuming the dark energy equation of state as $p_{\phi} = - \rho_{\phi}$.
The adiabatic regime naturally occurs without any additional constraint on scalar field equations.
The unified fluid description is valid for the background cosmology and for linear perturbations.
The equation of state of perturbations is the same as that of the background cosmology where all the effective results of the GCG paradigm are maintained.

Unfortunately, we cannot provide a sharp criterion on the potential and on the mass varying dependence on the scalar field to discriminate between these two possibilities: the GCG scenario or an effective unified fluid imitating the GCG via scalar fields driven by cosmon-{\em type} equations.
Many results for specific quintessence potentials that are found in the literature, when they reproduce stability and cosmic acceleration, are recovered, and speculative predictions for new scenarios featuring other mass dependencies on scalar fields can be made.
It remains open if the present approach can lead to a natural solution of the cosmological constant problem.
In the meanwhile we take the cosmon model analogy as an interesting phenomenological approach, through which we can reproduce the main characteristics of the GCG.

Given the fundamental nature of the underlying physics behind the Chaplygin gas and its generalizations,
it appears that it contains some of the key ingredients in the description of the Universe dynamics at early as well
as late times.
Our results suggest that the mass varying mechanism, when eventually derived from an exactly soluble field theory, which is noway trivial, can be the effective agent for the stability issue and for the cosmic acceleration of the universe, once it can effectively reproduce the main characteristics of a GCG scenario.
It also stimulates our subsequent investigation of the evolution of density perturbations, instabilities and the structure formation in such scenarios.
To summarize, we expect that the future precise data can provide more strong evidence to judge whether the dark energy is the cosmological constant and whether dark energy and dark matter can be unified into one cosmological background effective component.

\begin{acknowledgments}
We would like to thank for the financial support from the Brazilian Agencies FAPESP (grant 08/50671-0) and CNPq (grant 300627/2007-6).
\end{acknowledgments}

\end{document}